\documentclass[fleqn,twoside]{article}
\usepackage{espcrc2}

\usepackage{epsfig}
\usepackage[figuresright]{rotating}

\newcommand{\AmS}{{\protect\the\textfont2
  A\kern-.1667em\lower.5ex\hbox{M}\kern-.125emS}}

\newcommand{\order}[1]{{\mathcal O}(#1)}

\title{The charm quark mass with dynamical fermions}

\author{UKQCD Collaboration\\
        A.~Dougall\address{Department of Physics and Astronomy,
        University of Glasgow, Glasgow, G12 8QQ, UK},
        C.M.~Maynard\thanks{Presenter} \address{School of Physics,
        University of Edinburgh, Edinburgh EH9 3JZ, UK},
        C.~McNeile\address{Theoretical Physics, Dept. of Mathematical
        Sciences, University of Liverpool, Liverpool L69 3BX, UK}}
       
\begin{document}

\begin{abstract}
We compute the charm quark mass in lattice QCD and compare different
formulations of the heavy quark, and quenched data to that 
with dynamical sea quarks. We take the continuum limit of
the quenched data by extrapolating from three different lattice
spacings, and compare to data with two flavours of dynamical sea
quarks with a mass around the strange at the coarsest lattice 
spacing. Both the FNAL and ALPHA formalism are used.
We find the different heavy quark formulations have 
the same continuum limit in the quenched approximation, and limited
evidence that this approximation overestimates the charm quark mass.
\end{abstract}

\maketitle

\section{INTRODUCTION}
The charm quark mass is a relatively unknown quantity, yet this is
a quantity which is readily calculable on the lattice.  
The Particle Data Group~{\cite{Hagiwara:2002fs} quote 
\begin{equation}
\label{pdgCharmMass}
   1.0 < \overline{m}_{\rm charm}^{\overline{MS}}(\overline{m}_{\rm charm}) < 1.4 \ {\rm GeV}
\end{equation} 
Most lattice calculations to date have been done in the quenched
approximation. In this work we compare the results of the quenched
continuum limit from three lattice spacings, using different heavy
quark formulations to results with two flavours of dynamical fermions
with a mass around that of the strange quark.  An idea of how heavy
the sea quarks are can be seen by comparing the value of
$m_{PS}/m_V\sim 0.7$ to the physical value of
$m_\pi/m_\rho \sim 0.18$. Preliminary results were presented at the
previous lattice conference~\cite{Dougall:2003mx} and results for the
$D_s$ spectrum were presented in~\cite{Dougall:2003hv}.

\section{DETAILS OF THE CALCULATION}
We used the Wilson plaquette gauge action and the
Sheikholeslami-Wohlert action with a non-perturbative (NP)
determination of the coefficient $c_{SW}$. At each lattice spacing we
computed different definitions of the quark mass, for different values
of the hopping parameter. We fixed the charm mass using the $m_{Ds}$
pseudoscalar meson.  The details of the matching between the quenched
and dynamical ensembles can be found in \cite{Allton:1998gi}. The
details of the ensembles are shown in
Table~\ref{tab:ensembles}. Throughout this work the scale is set by
$r_0=0.5$ fm.

\begin{table} 
\caption{\label{tab:ensembles}Ensembles of gauge configurations.  
$\kappa_{\rm sea}=0$ denotes a quenched ensemble.
$N_G$ denote the  number of configurations.} 
\begin{tabular}{cccc} 
$(\beta,\kappa_{\rm sea})$ & Volume & $a^{-1}$ (GeV)& $N_G$\\
\hline
$(6.2,0)$&$24^3\times 48$ &$2.93$ & $216$ \\
$(6.0,0)$&$16^3\times 48$ &$2.12$ & $302$ \\
$(5.93,0)$&$16^3\times 32$ &$1.86$ & $278$ \\
$(5.2,0.1350)$&$16^3\times 32$ &$1.88$ & $395$ \\
\end{tabular}
\vspace{-10mm}
\end{table}

\subsection{Definitions of quark mass}
The quark mass can be defined in a number of ways. It can be defined
from the quark coupling, $\kappa$,
\begin{equation}
  am_0=\frac{1}{2}\left(\frac{1}{\kappa} - \frac{1}{\kappa_{\rm crit}}
  \right)
\label{eqn:m0}
\end{equation}
where $\kappa_{\rm crit}$ is the value of the hopping parameter which
corresponds to zero quark mass.  The renormalised quark mass, in the
mass independent scheme (ALPHA), is then
\begin{equation}
  am_V=Z_M\left(1+b_mam_0\right)am_0
  \label{eqn:mV}
\end{equation}

Another quark mass can be defined from the Axial Ward identity,
and correlation functions
\begin{equation}
   am_{ij} = \frac{ \sum_{\vec{x}}\langle \partial^4 A^4_{ij}(x)
   P_{ij}(0) \rangle } { 2 \sum_{\vec{x}}\langle P_{ij}(x) P_{ij}(0) \rangle }
\end{equation}
where $A$ is the improved Axial quark
current, $P$ the Pseudoscalar density, ans $i$ and $j$ label
quark flavour.
The subscripts $i,j$ denote
different quark flavours. For heavy-light correlation functions, the
renormalised quark mass is
\begin{eqnarray}
  am_A & = & \frac{Z_A}{Z_P}\left[ 2\left(1+(b_A-b_P)
    \frac{1}{2}a\overline{m}_0\right) \right.\\\nonumber
    &&\left. am_{HL} - \left(1+(b_A-b_P)am^L_0\right)am_{LL} \right]
  \label{eqn:mA}
\end{eqnarray}
where $\overline{m}_0 = \frac{1}{2}(m_0^H + m_0^L)$.

The values of all the improvement and renormalisation coefficients
have been determined to one loop in tadpole improved perturbation
theory~\cite{Bhattacharya:2000pn} in the $\overline{MS}$ scheme, as
the NP determinations are not available for the coarsest lattice
spacings. 

We also use the effective field theory formalism known as the FNAL
method. This presumes the dominance of ${\mathcal O}(am)^n$ lattice
artefacts over ${\mathcal O}(ap)^2$. This can be seen by considering
the distortion of the dispersion relation by the lattice,
\begin{equation}
E^2 = M_1^2 + \frac{M_1}{M_2}p^2 +  O(p^4)
\end{equation}
where $M_1$ is the rest mass and $M_2$ is the kinetic
mass, defined by 
\begin{equation}
\frac{1}{M_2} = \frac{\partial^2 E}{\partial p_k^2} \mid_{p=0}
\label{eq:KineticMass}
\end{equation}
It is the kinetic mass which is important for the dynamics of states
containing a heavy quark. The definitions of quark mass have $am_0$
corrections to all orders in $a$, to a particular order in $\alpha_S$.
At tree-level the quark mass is,
\begin{equation}
 am_1 = \log( 1 + a m_0)
\label{eq:m1}
\end{equation}
A perturbative definition of the kinetic mass is give by 
\begin{equation}
  am_2(am_1)= \frac{e^{am_1}\sinh(am_1)}{1+\sinh(am_1)}
\label{eq:m2}
\end{equation}
These quark masses can be used to determine the kinetic hadron mass,
\begin{equation}
aM_2^{PT} = aM_1 + (am_2 - am_1)
\end{equation}
The tree-level quark masses give a poor estimate of the kinetic mass
shift. However, using the one-loop masses, the perturbative $M_2$
tracks the NP $M_2$ from the dispersion relation, without the increase
in statistical errors coming from non-zero momentum states.  The
one-loop expressions relating the lattice quark mass and the pole
quark mass are,
\begin{equation}
  m_1^{(1)} = m_1^{(0)} + g^2 Z_{m_1}^{(1)} \tanh m_1^{(0)}
  \label{eqn:m1i}
\end{equation}
\begin{equation}
  m_2^{(1)} = m_2^{(0)}(m_1^{(0)} + g^2 m_1^{(0)}) ( 1 + g^2 Z_{m_2}^{(1)})
  \label{eqn:m2i}
\end{equation}
where the $Z$ functions are given in~\cite{Mertens:1998wx}. We now
refer to $m^{(1)}_{1,2}$ as $m_{1,2}$ for brevity. These
pole quark masses are then matched to the $\overline{MS}$ mass using
the one-loop expression from~\cite{Gray:1990yh}.

We interpolate each of the four quark masses $\{m_A,m_V,m_1,m_2\}$
from equations (\ref{eqn:mV}),(\ref{eqn:mA}),(\ref{eqn:m1i}) and
(\ref{eqn:m2i}) against the pseudoscalar heavy-light meson mass, where
the light quark mass has been interpolated to the strange quark
mass. We denote this $M_{Hs}$. All the quark masses have corrections
which scale with the quark mass, but the inverse hadron mass scales
with quark mass, so it is unclear whether to plot $am_Q$
vs. $1/M_{Hs}$ or vs. $M_{Hs}$. We do the latter as we are
interpolating in a finite range of $am_Q$, where a polynomial in
$am_Q$ can be expanded in terms of $1/am_Q$.  The heavy quark
interpolation for $m_A$ on the matched ensembles is shown in
figure~\ref{fig:HQInterpolation}. For ensembles with four different
heavy quarks ($\beta=\{6.2,6.0\}$) we use a quadratic function, for
ensembles with three ($\beta=\{5.93.5.2\}$ we use a linear function.

\begin{figure} 
\epsfig{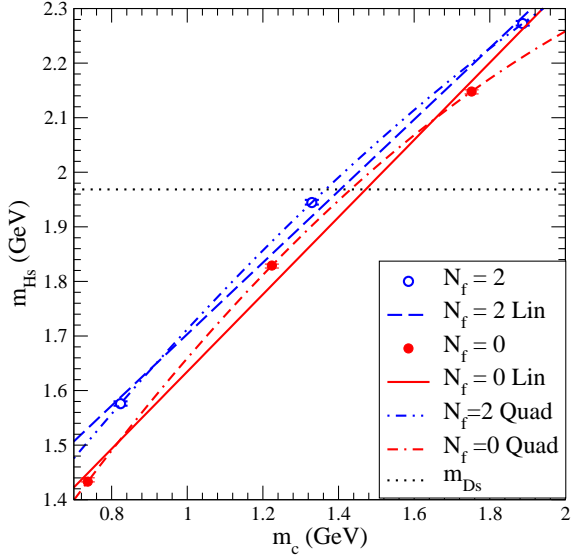} 
\vspace{-10mm} 
\caption{The hadron mass versus the $m_A$ quark mass for 
	the ensembles $(\beta=5.2,\kappa=0.1350)$ and 
	$\beta=5.93,\kappa=0)$.} 
\label{fig:HQInterpolation} 
\vspace{-10mm} 
\end{figure} 

The resulting charm mass is in the $\overline{MS}$ scheme at the scale
of the lattice spacing.  We evolve this mass using the renormalisation
group equation evaluated in a package called
RunDec~\cite{Chetyrkin:2000yt} to the scale invariant mass. We can
then compare the different quark masses and take the continuum limit
of the quenched data.

\begin{figure}
\epsfig{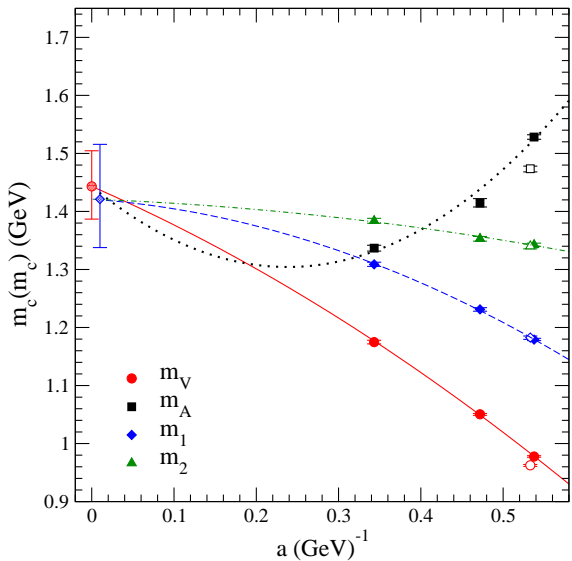}
\vspace{-10mm}
\caption{The quenched continuum limit of the scale invariant charm 
	quark mass. The closed (open) symbols show the quenched 
        (dynamical fermion) data. The shaded circle is the continuum 
	limit of the ALPHA masses and the diamond (offset for clarity)
	the FNAL masses.}
\label{fig:QCL}
\vspace{-8mm}
\end{figure}

\subsection{Comparing quark masses}
Shown in figure~\ref{fig:QCL} is the quenched continuum limit.  The
leading lattice artefacts are $\order{\alpha_S a m}$, $\order{\alpha_S
a}$ and $\order{a^2}$. For the masses $m_A$ and $m_V$ these are
clearly large and a linear extrapolation is not possible. However,
they should have a common continuum limit, as demonstrated by Rolf and
Sint~\cite{Rolf:2002gu}. We enforce the common continuum limit and
perform a simultaneous quadratic fit to both masses. Within large
errors this is consistent with the result from Rolf and Sint. We
perform a similar fit to the FNAL masses. The two methods
agree in the continuum limit.

The open symbols show the dynamical data. The dynamical ALPHA masses
are lower than the quenched data. However, this is not repeated for the
masses $m_1$ and $m_2$, and the sea quark mass is still rather heavy.
Moreover, the dynamical and quenched data could be effected by
different lattice artefacts.

Our analysis has larger statistical and systematic uncertainties than
the quenched result of Rolf and Sint, and other calculations. However,
we have demonstrated that the charm mass in the FNAL and ALPHA schemes
has the same continuum limit. We have also found evidence which
suggests the mass of the charm quark is too high in the quenched
approximation at fixed lattice spacing, even though the mass of the
sea quarks in the dynamical calculation is still rather heavy.

\end{document}